\documentclass[]{fairmeta}
\usepackage{amsmath,amsfonts}
\usepackage{algorithmic}
\usepackage{graphicx}
\usepackage{textcomp}
\usepackage{tikz}
\usepackage{tcolorbox}
\usepackage{enumitem}
\usepackage{listings}
\usepackage{booktabs} 
\usetikzlibrary{shapes.geometric, arrows, positioning}
\usepackage{siunitx}
\usepackage{hyperref}
\usepackage{tabularx}
\usepackage{ragged2e}
\usepackage{titlesec}

\newcommand{\sln}{Smartify}

\newcommand{\auditor}{{\it Auditor}}
\newcommand{\architect}{{\it Architect}}
\newcommand{\generator}{{\it Code Generator}}
\newcommand{\refiner}{{\it Refiner}}
\newcommand{\validator}{{\it Validator}}

\usepackage{hyperref}
\PassOptionsToPackage{hyphens}{url}
\usepackage{enumitem}
\setlist[itemize]{leftmargin=*} 
\usepackage{titlesec}
\titlespacing*{\section}{0pt}{1ex}{1ex} 
\titlespacing*{\subsection}{0pt}{0ex}{0ex} 

\title{Securing Smart Contract Languages with a Unified Agentic Framework for Vulnerability Repair in Solidity and Move
}
\author[1\dagger,3]{Rabimba Karanjai}
\author[2]{Lei Xu}
\author[1]{Weidong Shi}

\affiliation[1]{University Of Houston}
\affiliation[2]{{Kent State University}}
\contribution[\dagger]{rkaranjai@uh.edu}

\abstract{
The rapid growth of the blockchain ecosystem and the increasing value locked in smart contracts necessitate robust security measures. While languages like Solidity and Move aim to improve smart contract security, vulnerabilities persist. This paper presents \sln{}, a novel multi-agent framework leveraging Large Language Models (LLMs) to automatically detect and repair vulnerabilities in Solidity and Move smart contracts.  Unlike traditional methods that rely solely on vast pre-training datasets, Smartify employs a team of specialized agents working on different specially fine-tuned LLMs to analyze code based on the underlying programming concepts and language-specific security principles.  We evaluated \sln{} on a dataset for Solidity and a curated dataset for Move, demonstrating its effectiveness in fixing a wide range of vulnerabilities. 
  Our experimental results show that \sln{} (Gemma2+Codegemma) achieves state-of-the-art performance, surpassing existing LLMs and even enhancing the capabilities of general-purpose models, such as Llama 3.1. Notably, Smartify can incorporate language-specific knowledge, such as the nuances of Move, without requiring massive language-specific pretraining datasets. This work offers a detailed analysis of the performance of various LLMs on smart contract repair, highlighting the strengths of our multi-agent approach and providing a blueprint for developing more secure and reliable decentralized applications in the growing blockchain landscape. We also provide a detailed description to extend the proposed technology to other similar use cases.
}

\date{\today}

\begin{document}
\maketitle

\section{Introduction}

Smart contracts, self-executing agreements with terms directly written into code, have emerged as a cornerstone of blockchain technology~\cite{nath2014web,ray2023web3}. Their ability to automate transactions and eliminate intermediaries has led to widespread adoption in various sectors, including finance, supply chain management, and healthcare~\cite{zheng2018blockchain,karanjai2021conditional,kaleem2021event}. However, the increasing complexity of smart contracts has given rise to a growing concern: security vulnerabilities~\cite{vacca2021systematic}. These vulnerabilities, often stemming from coding errors or design flaws, can be exploited by malicious actors, leading to significant financial losses and damage to the reputation of blockchain projects. 

The financial implications of smart contract vulnerabilities are substantial. Reports indicate that cumulative losses from attacks against Ethereum smart contracts alone have exceeded USD 3.1 billion by 2023~\cite{li2023smart}. In the DeFi space, an estimated \$9.04 billion has been stolen due to vulnerabilities~\cite{wronka2023financial}. 
Notable incidents like the DAO hack of 2016, resulting in a \$55 million loss~\cite{popper2016hacking}, and the Poly Network hack in 2021, where over \$600 million was stolen~\cite{polyhack}, underscore the critical need for robust security measures. 

Traditional security auditing methods, although essential, often face limitations in terms of accuracy and scalability. This has spurred the exploration of automated techniques for vulnerability detection~\cite{10.1145/3238147.3238177,wang2020contractward}and repair, with Large Language Models (LLMs) emerging as a promising solution~\cite{joshi2023repair}. LLMs trained on vast datasets of code can learn to understand and generate code that adheres to specific programming paradigms and best practices. 
However, most of the tools available for smart contract safety are language-specific, especially \texttt{Solidity}~\cite{dannen2017solidity}.
For other languages, existing tools often require scanning of compiled bytecode~\cite{song2024empirical}.

% primarily relying on Solidity as the language of choice, as well as often sometimes requiring compiled bytecode for scanning for other languages~\cite{song2024empirical}.

Apart from \texttt{Solidity}, \texttt{Move}~\cite{blackshear2019move} has gained significant traction due to its strong focus on security. 
Its cutting-edge features, including a custom data type for secure operations, robust access controls via Move modules, and unique memory safety features~\cite{blackshear2022move}, have been
particularly noteworthy. Moreover, the Move Prover, a native security framework, provides an additional layer of protection \cite{dill2022fast}. Notably, several prominent blockchain platforms, such as Starcoin~\cite{starcoin}, Aptos~\cite{devaptos}, and Sui~\cite{blackshear2024sui}, have already adopted \texttt{Move}.

However, despite its promising architecture, the real-world security performance of \texttt{Move} modules remains largely untested. Unlike \texttt{Solidity}-based smart contracts, which have been extensively studied through empirical research and surveys, there is a scarcity of research focused specifically on Move modules. 
Although some methodologies have been proposed for identifying defects in \texttt{Move} modules or conducting formal verification~\cite{keilty2022model,park2024securing}, and empirical analysis~\cite{song2024empirical}, a significant knowledge gap persists. Specifically, large-scale investigations into the frequency of defects in real-world \texttt{Move} modules and identifying and repairing potential vulnerabilities are lacking, highlighting the need for further research in this area.

% This work proposes a novel framework for detecting and repairing vulnerabilities in smart contracts, focusing on the \texttt{Solidity} and \texttt{Move} languages from a programming language perspective. Our hypothesis relies on understanding the code and preventing known bad practices and unsafe code from being written before even compilation to prevent vulnerability. Our approach leverages the power of a multi-LLM agent system, combining the strengths of explanation and repair models. Our framework, \sln{}, leverages a multi-agent LLM framework to understand, critique, and repair code based on previously learned vulnerabilities as well as propose patches to repair them. 
% %
% By integrating an LLM specialized in code explanation with another focused on code repair, we aim to improve the accuracy and efficiency of the vulnerability remediation process.

This paper introduces Smartify, a framework that moves beyond naive LLM application by proposing a novel paradigm for structured, automated repair. We introduce Smartify not merely as a "multi-agent framework" but as a \textbf{structured, collaborative system that mimics the workflow of an expert human security audit team}. This contrasts with existing LLM-powered tools like \texttt{ContractTinker}, which utilize a linear Chain-of-Thought (CoT) reasoning process to break down the repair task. Smartify's novelty lies in its process-centric design, where specialized agents (Auditor, Architect, Code Generator, Refiner, Validator) assume distinct roles within a delegative and iterative workflow. The Architect agent, for instance, does not simply reason about the next step; it formulates a comprehensive repair strategy that is then executed by other agents. Crucially, the Refiner-Validator duo establishes a feedback loop for iterative quality assurance, a feature essential for generating trustworthy patches.

In this paper, our aim is to answer the following research questions related to software engineering using AI agents and the landscape of complex smart contract reasoning.
\begin{itemize}
    \item \textbf{RQ1:} \textit{Do the present state-of-the-art LLMs can explain a Smart Contract code correctly?}
    \item \textbf{RQ2:} \textit{Can they detect and explain bad coding practices or specific mistakes leading to bugs or vulnerabilities in a smart contract code?}
    \item \textbf{RQ3:} \textit{Can we encode programming language-specific knowledge to train the LLMs to understand unsafe and buggy codes in detail enough to repair them?}
    \item \textbf{RQ4:} \textit{Can LLMs repair the bugs and fix the vulnerability?}
    \item \textbf{RQ5:} \textit{Does the proposed post-training framework be generalizable to a larger set of pre-trained LLMs?}
\end{itemize}

\underline{The main contributions of our work are the following:}
\begin{itemize}[leftmargin=*]
    \item We introduce a novel \textbf{role-based multi-agent architecture} for smart contract repair that models the collaborative workflow of a security audit team, enhancing structured reasoning over monolithic LLM approaches.
    \item We propose a rigorous methodology combining \textbf{specialized fine-tuned models for deep vulnerability analysis}, \textbf{Retrieval-Augmented Generation (RAG) for language-specific context}, and an \textbf{iterative self-refinement loop} for ensuring patch quality.
    \item We conduct the first \textbf{multi-faceted empirical evaluation} of an LLM-based repair tool for both Solidity and Move, using metrics that assess not only correctness but also \textbf{exploit mitigation effectiveness}, \textbf{semantic preservation}, and \textbf{code quality}.
    \item We present a comprehensive \textbf{ablation study} that systematically dissects our framework to quantify the distinct contribution of each architectural component to the overall repair performance.
\end{itemize}

\section{Related Work}

This section provides a critical review of the existing landscape in smart contract security, positioning Smartify relative to traditional analysis techniques and emerging LLM-based repair methodologies.

\subsection{Traditional Smart Contract Security Analysis}
Automated security analysis for smart contracts has traditionally been dominated by static, dynamic, and formal verification techniques. Each approach offers distinct advantages but also possesses inherent limitations, particularly when confronted with complex, logic-based vulnerabilities.

\subsection{Smart Contract Security Auditing}

Various tools and techniques have been developed for detecting vulnerabilities in smart contracts:

\textbf{Static Analysis Tools:} Tools like Mythril~\cite{muellerfile} and Slither~\cite{feist2019slither} analyze contract source code to identify potential vulnerabilities. They perform symbolic execution and taint analysis to detect patterns associated with common vulnerabilities.

\textbf{Dynamic Analysis Tools:} Tools like Manticore~\cite{mossberg2019manticore} and Echidna~\cite{grieco2020echidna} execute contracts with various inputs to uncover runtime errors. They use fuzzing and symbolic execution techniques to explore different execution paths and identify potential issues.

\textbf{Formal Verification:} This approach uses mathematical techniques to rigorously prove the correctness of a contract's code against a formal specification. Tools like KEVM~\cite{hildenbrandt2018kevm} and CertiK's DeepSEA have been developed for formal verification of smart contracts~\cite{zhong2020move}.

While these tools are valuable, they often have limitations in accuracy, scalability, and the ability to handle the complexities of real-world smart contracts.

\subsection{LLM-Powered Smart Contract Repair: A Taxonomy}
The application of LLMs to automated program repair is a rapidly advancing field, with several distinct methodologies emerging.

\subsubsection{Monolithic LLM Approaches}
Early efforts in this domain involved using large, general-purpose LLMs like GPT-3 with intricate, zero-shot or few-shot prompts to generate patches. While demonstrating feasibility, these approaches often lack the domain-specific knowledge required for the high-stakes environment of smart contracts. They are prone to generating syntactically correct but semantically flawed or insecure code, as they lack a deep, ingrained understanding of blockchain-specific security paradigms \cite{blackshear2022move, bobadilla2025automated}.

\subsubsection{Chain-of-Thought (CoT) and Static Analysis Integration}
To address the limitations of monolithic models, more structured approaches have been developed. A prominent example is \textbf{ContractTinker}, a tool designed for real-world smart contract repair \cite{wang2024contracttinker}. \texttt{ContractTinker} employs a Chain-of-Thought (CoT) mechanism to guide an LLM through a sequence of reasoning steps: vulnerability localization, analysis, and patch generation. To ground the LLM's reasoning and mitigate hallucination, it integrates static analysis techniques, including dependency analysis and program slicing, to provide relevant context from audit reports and the source code itself.

While \texttt{ContractTinker} represents a significant advancement by imposing a logical structure on the repair process, its CoT approach remains a fundamentally linear and sequential reasoning pipeline. In contrast, Smartify's architecture is \textbf{delegative and iterative}. The Architect agent formulates a comprehensive, high-level plan, which is then delegated to specialized agents for execution. Furthermore, the explicit Refiner-Validator loop introduces a crucial mechanism for feedback and iterative quality improvement, a feature not explicitly detailed in the \texttt{ContractTinker} workflow. This architectural distinction moves beyond a simple chain of thought to a collaborative problem-solving process.

\subsubsection{Multi-Agent Systems for Code Tasks}
The concept of using multiple LLM-based agents to collaborate on complex tasks has gained traction in the broader software engineering domain, with applications in areas like code translation and generation \cite{huang2024opencoder, jiang2018contractfuzzer}. These systems leverage the principle of specialization, assigning different roles or sub-tasks to individual agents to achieve a more robust and accurate outcome than a single agent could.

While Smartify aligns with this general trend, its novelty lies in its \textbf{domain-specific agent roles tailored explicitly for the vulnerability repair workflow}. Instead of a generic "translator" or "coder" agent, Smartify's agents embody the distinct functions of a human security audit team: the Auditor for analysis, the Architect for strategic planning, the Code Generator for implementation, and the Refiner/Validator for quality assurance. This specialization allows for a more nuanced and effective approach to the highly specific and critical task of securing smart contracts.

Our proposed framework, \textbf{Smartify}, addresses these challenges by combining the strengths of specialized LLMs within a multi-agent architecture. It leverages language-specific fine-tuning, safety classifiers, and Retrieval-Augmented Generation (RAG) to enhance the accuracy and security of generated code repairs. 
% Furthermore, \sln{} incorporates the SolMover~\cite{kara} tool to facilitate cross-language translation between Solidity and Move, expanding its applicability within the blockchain ecosystem.

In the following sections, we detail the architecture of \sln{}, describe the experimental setup, present the evaluation results, and discuss the implications of our findings for the future of smart contract security.

\section{Data Collection and Analysis Methodology}

This research employs a multi-faceted approach to investigate the security of smart contracts, focusing on both \texttt{Solidity} and \texttt{Move} programming languages. The methodology encompasses collecting and analyzing two distinct datasets: \texttt{Solidity}-based, \texttt{Move}-based source code. Each dataset serves a specific purpose in addressing the research questions and contributing to a comprehensive understanding of smart contract vulnerabilities.

\subsection{Importance of Dataset Categorization}

For several reasons, categorizing the datasets based on programming language (\texttt{Solidity} and \texttt{Move}) and code representation is crucial. It allows for a focused analysis of language-specific vulnerabilities and coding practices. As a more mature language, \texttt{Solidity} exhibits a different vulnerability landscape than the newer \texttt{Move} language. Examining them separately enables the identification of unique challenges and security considerations associated with each language. 
% Second, the distinction between source code and bytecode datasets facilitates different types of analysis. 

% Source code analysis permits manual auditing, formal verification, and static analysis, while bytecode analysis is essential for studying deployed contracts and conducting large-scale assessments of on-chain security. This layered approach provides a holistic view of the smart contract security ecosystem, covering development and deployment phases.

\subsection{Dataset Descriptions}

\subsubsection{\texttt{Solidity}-based Dataset}
This dataset comprises a collection of vulnerable \texttt{Solidity} smart contracts sourced from the "Not-So-Smart Contracts" repository curated by Trail of Bits \cite{githubGitHubCryticnotsosmartcontracts}. This repository is renowned for its comprehensive set of contracts that intentionally exhibit a variety of common vulnerabilities. These vulnerabilities were chosen for inclusion because of their prevalence in real-world decentralized applications and their representation of typical errors during smart contract development. 
The dataset contains 60 vulnerable contracts, encompassing 8 distinct vulnerability categories. Table \ref{tab:solidity-distribution} shows these categories' distribution.

% follows: 25\% Reentrancy, 16.7\% Integer Overflow/Underflow, 13.3\% Denial of Service (DoS), 20\% Access Control Issues, 8.3\% Uninitialized Storage Pointers, 5\% Tx.origin Misuse, 6.7\% Timestamp Dependency, and 5\% Gas Limit and Out-of-Gas Vulnerabilities.

%\begin{table}[ht]
\begin{table}[!htbp]
\centering
\caption{Distribution of Vulnerabilities in the \texttt{Solidity} Dataset.}
\label{tab:solidity-distribution}
%\footnotesize
\scriptsize
\begin{tabularx}{\columnwidth}{|>{\raggedright\arraybackslash}X|S[table-format=2.0]|S[table-format=2.1]|}
\hline
\textbf{Vulnerability Type} & {\textbf{Number of Contracts}} & {\textbf{Percentage (\%)}} \\ \hline
Reentrancy & 15 & 25.0 \\ \hline
Integer Overflow/Underflow & 10 & 16.7 \\ \hline
Denial of Service (DoS) & 8 & 13.3 \\ \hline
Access Control Issues & 12 & 20.0 \\ \hline
Uninitialized Storage Pointers & 5 & 8.3 \\ \hline
Tx.origin Misuse & 3 & 5.0 \\ \hline
Timestamp Dependency & 4 & 6.7 \\ \hline
Gas Limit and Out-of-Gas Vulnerabilities & 3 & 5.0 \\ \hline
\textbf{Total} & \textbf{60} & \textbf{100.0} \\ \hline
\end{tabularx}
\end{table}

% \begin{figure}[ht]
% \centering
% \begin{tikzpicture}
% \begin{axis}[
%     axis x line=none, 
%     axis y line=none,
%     enlarge x limits=0.15,
%     enlarge y limits=0.15,
%     ybar,
%     ymin=0,
%     bar width=15pt,
%     nodes near coords,
%     nodes near coords align={vertical},
%     symbolic x coords={Reentrancy,Integer Overflow/Underflow,Denial of Service (DoS),Access Control Issues,Uninitialized Storage Pointers,Tx.origin Misuse,Timestamp Dependency,Gas Limit and Out-of-Gas Vulnerabilities},
%     xtick=data,
%     xticklabel style={rotate=45,anchor=east},
%     ylabel={Percentage (\%)},
%     title={Distribution of Vulnerabilities in the Solidity Dataset},
% ]
% \addplot coordinates {
%     (Reentrancy,25)
%     (Integer Overflow/Underflow,16.7)
%     (Denial of Service (DoS),13.3)
%     (Access Control Issues,20)
%     (Uninitialized Storage Pointers,8.3)
%     (Tx.origin Misuse,5)
%     (Timestamp Dependency,6.7)
%     (Gas Limit and Out-of-Gas Vulnerabilities,5)
% };
% \end{axis}
% \end{tikzpicture}
% \caption{Distribution of Vulnerabilities in the Solidity Dataset (Bar Chart)}
% \label{fig:solidity-bar-chart}
% \end{figure}

\subsubsection{\texttt{Move}-Based Dataset (Source Code)}
This dataset encompasses the source code of 92 real-world \texttt{Move} projects, comprising 652 individual modules. These projects were part of Aptos \cite{devaptos}, Sui \cite{blackshear2024sui}, and Starcoin \cite{starcoin}. These projects span various application domains, as depicted in Table \ref{tab:move-source-distribution}.
% including 41 Decentralized Finance (DeFi) projects, 22 Token projects, 18 Bridge projects, 3 Library projects, 3 Infrastructure projects, and five miscellaneous projects. 
%
The total number of \texttt{Move} projects is 92, and the total number of \texttt{Move} modules within these projects is 652.

\begin{table}[!htbp]
\centering
\caption{Distribution of \texttt{Move} Projects by Application Domain.}
\label{tab:move-source-distribution}
%\footnotesize
\scriptsize
\begin{tabularx}{\columnwidth}{|>{\raggedright\arraybackslash}X|S[table-format=2.0]|S[table-format=2.1]|}
\hline
\textbf{Application Domain} & {\textbf{Number of Projects}} & {\textbf{Percentage (\%)}} \\ \hline
Decentralized Finance & 41 & 44.6 \\ \hline
Token & 22 & 23.9 \\ \hline
Bridge & 18 & 19.6 \\ \hline
Library & 3 & 3.3 \\ \hline
Infrastructure & 3 & 3.3 \\ \hline
Other & 5 & 5.4 \\ \hline
\textbf{Total} & \textbf{92} & \textbf{100.0} \\ \hline
\end{tabularx}
\end{table}
% \subsubsection{Move-Based Dataset (Bytecode)}

% This dataset consists of 37,302 bytecode instances of deployed Move modules collected from Aptos and Sui blockchains. The dataset is divided into 15,479 bytecode instances from Aptos and 21,823 from Sui. Starcoin was excluded due to its significantly lower transaction volume, representing only a negligible fraction of Aptos and Sui's activity.  The bytecode instances from Aptos constitute approximately 41.5\% of the total, while those from Sui make up the remaining 58.5\%. This distribution is visualized in Figure \ref{fig:move-bytecode-pie-chart}.

For both the \texttt{Move}-based datasets, we utilize Song et al.'s~\cite{song2024empirical} work to compare the vulnerability detection part. While the prior work is directed towards detection, the same dataset helps us compare \sln{}'s performance on both detection and repair.

\subsection{Evaluation of \sln{}} \label{evaluationmethod}

\sln{} is designed with two core functionalities: detecting and repairing unsafe coding patterns in smart contracts. We utilize the previously described datasets to evaluate these capabilities rigorously, encompassing both \texttt{Solidity} and \texttt{Move} code. The evaluation process focuses on the complete output of \sln{} rather than individual components, reflecting its nature as an integrated solution for smart contract security. Performance is measured using the Pass@1 score.

% \textit{Pass@1} is a metric that quantifies the percentage of problems for which a correct solution is generated on the first attempt. In the context of code generation and repair, it signifies the system's ability to produce functional and secure code without requiring iterative refinement or manual intervention. A higher Pass@1 score indicates greater accuracy and efficiency. Pass@1 is particularly important for this research as it reflects the practical applicability of \sln{} in real-world scenarios. A high Pass@1 score suggests that \sln{} can be effectively used to automatically detect and repair vulnerabilities in smart contracts, reducing the burden on developers and enhancing the overall security of blockchain applications.

\subsection{Agent-Based Code Repair Process for Smart Contracts}

Our approach leverages a multi-agent system inspired by established software development methodologies but tailored explicitly for the automated repair of \texttt{Solidity} and \texttt{Move} smart contracts. This system employs five specialized agents: an \auditor{}, an \architect{}, a \generator{}, a 
\refiner{}, and a \validator{}. The process incorporates a self-refinement loop and a final validation step, ensuring high accuracy and security. Each agent plays a distinct role in a structured workflow, detailed below.

\begin{figure}
    \centering
    \includegraphics[width=0.8\linewidth,height=5.8cm]{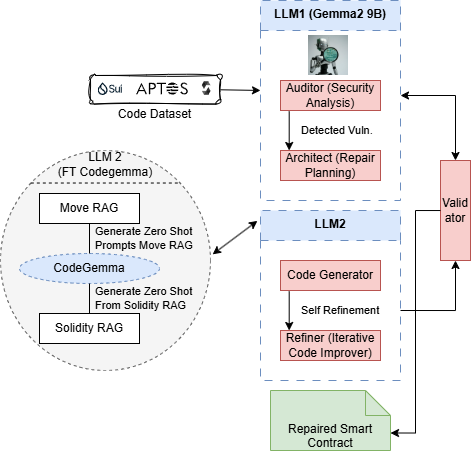}
    \caption{Agentic architecture of \sln{}.}
    \label{fig:archi}
    \vspace{-0.3cm}
\end{figure}

%\subsubsection{Agent Roles and Responsibilities}

\begin{itemize}[leftmargin=*]
    \item \textbf{Auditor:} This agent is the cornerstone of security analysis. It is fine-tuned on a comprehensive corpus of \texttt{Solidity} and \texttt{Move} code documentation, encompassing syntax, semantics, and best practices. Furthermore, it is safety-aligned using a classifier adapted from Google's Responsible AI toolkit. This classifier has been meticulously modified to enforce language-specific rules and safe coding practices, effectively preventing the generation of unsafe or unsupported code constructs.  

    This alignment is of paramount importance. For \texttt{Move}, it ensures that the generated code strictly adheres to the conventions of the target blockchain (e.g., Sui or Aptos). It prevents the accidental introduction of elements from one Move variant into another or the inclusion of unsupported Rust paradigms. This is crucial because Move, derived from Rust, has unique features and limitations. For \texttt{Solidity}, it enforces established security best practices and prevents the generation of code patterns known to be vulnerable.

    The Auditor's primary responsibility is to meticulously scan the input smart contract code (either \texttt{Solidity} or \texttt{Move}) to identify potential vulnerabilities and unsafe patterns. Its secondary, yet vital, role is to serve as the final \validator{} of the repaired code.
    \item \textbf{Architect:} This agent receives the output from the \auditor{}, which includes a detailed report of identified vulnerabilities and unsafe code segments. The \architect{}'s role is to devise a high-level strategic plan for addressing these issues. This plan does not involve generating code directly. Instead, it outlines the necessary modifications, refactoring, and improvements to rectify the identified problems. This plan is a comprehensive blueprint for the \generator{}, guiding the code repair process.
    \item \textbf{Code Generator:} This agent is a general-purpose code LLM. Its strength lies in its ability to leverage Retrieval-Augmented Generation (RAG) from two distinct data stores, one dedicated to \texttt{Solidity} and the other to \texttt{Move}. 
    These data stores contain a collection of best practices and relevant documentation for the respective programming languages. 

    Using the \architect{}'s plan as a guide, the Code Generator selects and adapts relevant examples from the appropriate RAG datastore. This dynamic, context-aware retrieval of few-shot examples significantly enhances the \generator{}'s ability to produce accurate and secure code repairs. It ensures the generated code adheres to language-specific conventions and incorporates established best practices.
    \item \textbf{Refiner:} This agent's role is to enhance the code quality produced by the \generator{}. It achieves this through iterative self-refinement, essentially acting as its critic. The \refiner{} uses the same underlying LLM as the \generator{} but with a different prompt that focuses on improving the code quality based on best practices and potential improvements that it might detect from a higher level.
    \item \textbf{Validator:} This agent acts as a final checkpoint. After the refinement stage, it re-employs the \auditor{} agent to re-evaluate the code. The \validator{}'s objective is to ensure that all the previously identified vulnerabilities have been adequately addressed and that no new vulnerabilities have been introduced during the repair and refinement process.
\end{itemize}

\section{\sln{} System Architecture and Workflow}

\sln{} operates through a five-agent system designed for automated intelligent contract vulnerability detection and repair. The system functions as shown in Figure~\ref{fig:archi}.

The Smartify system operates in a five-phase process to automatically repair smart contract code.  Firstly, in the Input \& Initial Audit phase, the smart contract code, written in either \texttt{Solidity} or \texttt{Move}, is fed into the system. The \auditor{}, an LLM based on Gemma2 9B, analyzes the code to detect potential vulnerabilities and produces a report detailing its findings.  Secondly, during Repair Planning, the \architect{} receives this vulnerability report and formulates a high-level repair plan that outlines the necessary code modifications to address the identified issues. Thirdly, in Code Generation \& Refinement, an LLM called CodeGemma, which has been fine-tuned for code generation and is equipped with Retrieval-Augmented Generation (RAG) capabilities, takes the lead. It utilizes separate \texttt{Move} RAG and \texttt{Solidity} RAG components to provide language-specific context. The Code Generator, part of CodeGemma, uses the repair plan to generate the modified code, selecting the appropriate RAG based on the input language and able to perform \texttt{Solidity} to \texttt{Move} translation when necessary. 

Subsequently, a Self-refinement process is initiated, and the \refiner{} component iteratively improves the generated code's quality, readability, and efficiency.  Fourthly, in the Validation phase, the \validator{} (the same agent as the \auditor{}) performs a final security audit on the refined code to ensure all identified vulnerabilities have been resolved. Finally, the system outputs the repaired smart contract code.

% \begin{algorithm}[H]
% \caption{Smartify: Automated Smart Contract Repair}
% \label{alg:smartify}
% \begin{algorithmic}[1]
% \STATE \textbf{Input:} Smart contract code $C$ (Solidity or Move)
% \STATE \textbf{Output:} Repaired smart contract code $C'$

% \STATE \textbf{Procedure} Smartify($C$)
%     \STATE // \textbf{Phase 1: Initial Audit}
%     \STATE $V \gets $ Auditor($C$) \COMMENT{Auditor: LLM1 (Gemma2 9B)} 
%     \STATE \COMMENT{$V$: List of detected vulnerabilities}
    
%     \STATE // \textbf{Phase 2: Repair Planning}
%     \STATE $P \gets $ Architect($V$) \COMMENT{Architect}
%     \STATE \COMMENT{$P$: High-level repair plan}

%     \STATE // \textbf{Phase 3: Code Generation and Refinement}
%     \STATE $L \gets $ DetermineLanguage($C$) \COMMENT{Determine if $C$ is Solidity or Move}
%     \IF{$L$ = Solidity}
%       \STATE $R \gets $ SolidityRAG()
%     \ELSE
%       \STATE $R \gets $ MoveRAG()
%     \ENDIF
    
%     \STATE $C_{gen} \gets $ CodeGenerator($P, R$) \COMMENT{CodeGenerator: LLM2 (FT CodeGemma), utilizes RAG}
%     \STATE $C' \gets $ Refiner($C_{gen}$) \COMMENT{Refiner: Iterative code improvement}
    
%     \STATE // \textbf{Phase 4: Validation}
%     \STATE $V' \gets $ Validator($C'$) \COMMENT{Validator: same agent as Auditor}
%     \IF{$V' = \emptyset$} \COMMENT{No vulnerabilities detected}
%         \STATE \textbf{return} $C'$
%     \ELSE
%         \STATE \textbf{goto} Step 4 \COMMENT{Repeat the process if new vulnerabilities are found}
%     \ENDIF
% \end{algorithmic}
% \end{algorithm}

The process may iterate to step 3 or 4 if the \validator{} identifies any issues. Each step is vital in ensuring the smart contract code's accurate and secure repair. The workflow is designed to be efficient and effective, leveraging each agent's strengths to achieve the desired outcome.

\subsection{Agent Prompting Strategy}

The agents within \sln{} are driven by carefully crafted prompts that guide their actions and ensure consistent performance. We employ a standardized prompt template adapted from established practices in LLM-based agent systems. The template is structured as follows:

\begin{tcolorbox}[
  colback=gray!10, % Light gray background
  colframe=gray!50, % Medium gray border
  title=Prompt Template,
  fonttitle=\bfseries,
  size=small,
  boxrule=0.75mm, % Thicker border for emphasis
  rounded corners, % Smooth corners for aesthetics
  left=1mm, % Small padding on the left
  right=1mm, % Small padding on the right
  top=1mm, % Small padding on the top
  bottom=1mm, % Small padding on the bottom
  label=box:prompt template
]
{\footnotesize
\textbf{Role:} You are a \texttt{[role]} specializing in \texttt{[Solidity/Move]} smart contracts.

\textbf{Task:} \texttt{[task]}

\textbf{Instruction:} Based on the provided Context, please follow these steps: \texttt{[numbered steps]}

\textbf{Context:} ...
}
\end{tcolorbox}

This template is broken down into the following components.
% \begin{itemize}
%     \item \textbf{Role:} Specifies the agent's role (e.g., Auditor, Architect, Code Generator, Refiner, Validator).
%     \item \textbf{Task:} Describes the specific task the agent is expected to perform (e.g., "Identify vulnerabilities," "Generate repair plan," "Generate code").
%     \item \textbf{Instruction:} Provides a detailed, step-by-step guide on how to accomplish the task. This section leverages chain-of-thought reasoning to guide the agent's actions and decision-making process.
%     \item \textbf{Context:} Contains all the necessary information for the agent to perform its task. This may include the input code, audit reports, architectural plans, code examples from the RAG datastore, and the conversation history between agents.
% \end{itemize}

Each agent in our framework is defined by four key components: the Role, which designates the agent’s specific function (such as Auditor, Architect, or Code Generator); the Task, which outlines the agent’s particular objectives; the Instruction, which provides detailed step-by-step guidance using chain-of-thought reasoning; and the Context, which encompasses all necessary information including input code, audit reports, architectural plans, RAG datastore examples, and inter-agent conversation history.

Table \ref{tab:agent-prompts} shows how this template is adapted for each agent.

\begin{table}[ht]
\centering
\caption{Agent Prompts for Smart Contract Repair.}
\label{tab:agent-prompts}
\footnotesize
\renewcommand{\arraystretch}{1.2} % Adjust row height for readability
\setlength{\tabcolsep}{4pt} % Adjust column spacing
\begin{tabularx}{\linewidth}{|p{1.5cm}|X|X|p{2.0cm}|}
\hline
\textbf{Role} & \textbf{Task} & \textbf{Instruction} & \textbf{Context} \\ \hline
{\bf Auditor} & Identify vulnerabilities and unsafe patterns in Solidity/Move code. & Analyze the code for security vulnerabilities and generate a detailed report. & Input smart contract code (Solidity/Move). \\ \hline
{\bf Architect} & Create a high-level plan to address vulnerabilities identified by the Auditor. & Review the Auditor's report and develop a plan outlining necessary modifications. & Auditor's report. \\ \hline
{\bf Code Generator} & Generate Repaired Solidity/Move code based on the Architect's plan and RAG examples. & Consult the Architect's plan, retrieve examples from the RAG datastore, and generate repaired code. & Architect's plan, Solidity/Move code examples from RAG. \\ \hline
{\bf Refiner} & Iteratively refine the generated code to improve quality and efficiency. & Review the generated code, identify areas for improvement, and refine accordingly. & Generated code, previous iteration code (if any). \\ \hline
{\bf Validator} & Perform a final security check on the repaired code. & Analyze the repaired code for vulnerabilities, verify issue resolution, and ensure no new vulnerabilities. & Repaired smart contract code. \\ \hline
\end{tabularx}
\end{table}

\subsection{Hardware and Model Fine-tuning}

The development and deployment of \sln{} leveraged a heterogeneous computing environment, utilizing high-performance GPUs for computationally intensive tasks and a more resource-efficient setup for inference.

\subsubsection{Fine-tuning Setup}

\begin{itemize}[leftmargin=*]
    \item \textbf{Hardware:} Fine-tuning leveraged a cluster of \textbf{four NVIDIA A100 GPUs} for computationally demanding pattern learning in \texttt{Solidity} and \texttt{Move} code.
    \item \textbf{Model:} Based on the \textbf{Gemma 9B model}, selected for strong code-related task performance and fine-tuning adaptability, particularly in instruction following. Fine-tuned on a dataset of \texttt{Solidity} and \texttt{Move} code, vulnerability examples, best practices, and documentation, augmented with outputs from earlier pipeline stages to enhance safety issue detection.
    \item \textbf{Training Recipe:} Supervised learning paradigm. Trained to predict correct outputs (e.g., vulnerability reports, safe code patterns) from inputs (e.g., \texttt{Solidity}/\texttt{Move} code, vulnerability descriptions).
    \begin{itemize}[leftmargin=*]
        \item \textbf{Data Pre-processing:} Tokenization, normalization, and input-output pair creation ensured data consistency and quality.
        \item \textbf{Hyperparameter Optimization:} Learning rate (1e-5), batch size (8, due to memory constraints), and training epochs (5, as validation loss plateaued) optimized via grid search and manual tuning.
        \item \textbf{Regularization:} Dropout and weight decay used to prevent overfitting and improve generalization.
        \item \textbf{Evaluation Metrics:} Accuracy, precision, recall, and F1-score on a held-out validation set monitored model performance.
    \end{itemize}
\end{itemize}

\subsubsection{Inference Setup}

\begin{itemize}[leftmargin=*]
    \item \textbf{Hardware:} Inference was performed on a single \textbf{NVIDIA RTX 4090 GPU}, balancing performance and cost-effectiveness for real-time code repair.
    \item \textbf{Models:}
    \begin{itemize}
        \item \textbf{Code Generator and Refiner:} These agents utilize a fine-tuned \textbf{CodeGemma} model, initially pre-trained on a limited \texttt{Move} corpus and further instruction-tuned to follow Architect-generated "recipe" patterns. Fine-tuning on Architect outputs ensured it understood these instructions, and pre-training on a limited \texttt{Move} corpus ensured basic syntax understanding.
        \item \textbf{Comparison Model:} A stock \textbf{Llama 3.1} model was used in some experiments for comparative analysis, helping assess the gains from fine-tuning and instruction tuning.
    \end{itemize}
\end{itemize}

\subsubsection{Key Considerations}

\begin{itemize}[leftmargin=*]
    \item A balance between performance requirements, resource availability, and cost considerations drove the choice of hardware and models.
    \item The fine-tuning process for the \auditor{} was particularly resource-intensive due to the complexity of the task and the size of the model.
    \item The use of a smaller, more efficient GPU for inference makes the system more accessible for practical deployment.
    \item The comparison with a stock Llama 3 model provides valuable insights into the effectiveness of our fine-tuning and instruction-tuning strategies.
\end{itemize}

This heterogeneous setup, combining high-performance GPUs for training and a more efficient GPU for inference, allows \sln{} to effectively address the computational demands of both model development and deployment. The detailed description of the fine-tuning process provides transparency and allows for replication of our results.
\section{Experimental Results and Discussion}

We run our experiments as defined in Section \ref{evaluationmethod}. We report the results as well as the empirical performance of our models. Through that, we will try to answer our Research Questions individually in this section.

Along with \sln{}, we have run the benchmark for the following models.

% granite-code\:8b-instruct, codegemma\:7b-instruct, deepseek\-coder\-v2, starcoder2, codegeex4, codestral, deepseek-coder\:33b, codellama\:13b, codeqwen\:7b\-chat\-v1.5\-q8\_0, qwen2.5\-coder, gemma2, gemma2:27b, llama3.2, opencoder:8b-instruct-fp16, llama3.3.
\begin{table}[!htbp]
    %\scriptsize % Reduce font size for the table
    \centering
    \setlength{\tabcolsep}{2pt} % Adjust column spacing
    \renewcommand{\arraystretch}{1.2} % Adjust row spacing
    \caption{Comparison of Code and Non-Code Models.}
    \label{tab:model-comparison}
    %\footnotesize
    \scriptsize
    \begin{tabular}{@{}p{2.5cm}p{1.5cm}p{1.5cm}p{1cm}@{}}
        \toprule
        \textbf{Model Name} & \textbf{Parameters} & \textbf{Quantization} & \textbf{Code Model} \\ 
        \midrule
        Granite-Code     & 8B    & FP16 & Yes \\
        CodeGemma        & 7B    & FP16 & Yes \\
        DeepSeek-Coder-v2            & N/A   & N/A  & Yes \\
        StarCoder2                   & 15B   & FP16 & Yes \\
        CodeGeex4                    & 13B   & N/A  & Yes \\
        CodeStral                    & 7B    & FP16 & Yes \\
        DeepSeek-Coder           & 33B   & N/A  & Yes \\
        CodeLlama~\cite{roziere2023code}                & 13B   & N/A  & Yes \\
        CodeQwen  & 7B    & Q8\_0 & Yes \\
        Qwen2.5-coder                & 2.5B  & N/A  & Yes \\
        Gemma2                       & N/A   & N/A  & Yes \\
        Gemma2:27b                   & 27B   & FP16 & Yes \\
        Llama3.2                     & 3.2B  & FP16 & No  \\
        OpenCoder   & 8B    & FP16 & Yes \\
        Llama3.3                     & 3.3B  & FP16 & No  \\
        \bottomrule
    \end{tabular}
\end{table}

The models were chosen according to the top 8 models at Hugging Face Big Code Leaderboard~\cite{huggingfaceCodeModels} at the time of this work, and also adding general-purpose models, which are supposed to be better at reasoning.

\subsection{Solidity}

\begin{figure}
    \centering
    \includegraphics[width=1\linewidth,height=3.6cm]{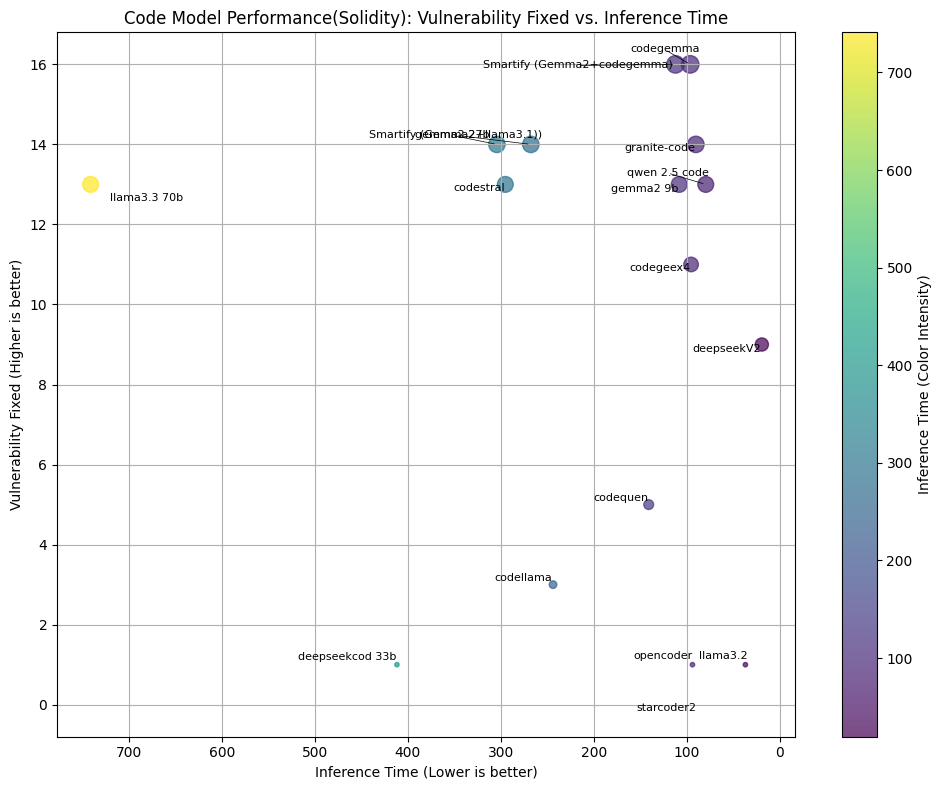}
    \caption{Code Repair: Solidity.}
    \label{fig:solidityrepair}
\end{figure}

This section presents the evaluation results of various code generation models on repairing vulnerabilities in Solidity smart contracts, specifically focusing on the ``Not So Smart Contracts'' dataset from the Trail of Bits GitHub repository. This dataset is a collection of intentionally vulnerable Solidity contracts designed to test the ability of automated tools to detect and repair common security flaws. It contains diverse vulnerabilities, including re-entrancy, integer overflow/underflow, access control issues, and timestamp dependence, among others. The dataset has been publicly available for a significant period, raising the possibility that some or all of its contents might be present in the pre-training data of the evaluated models. We analyze the performance of these models based on two key metrics: the number of vulnerabilities fixed and the average inference time, as summarized in Table \ref{tab:model_performance} and Figure \ref{fig:solidityrepair}. We also introduce our framework, \sln{}, and demonstrate its effectiveness in enhancing model performance.

\begin{table}[!htbp]
  \caption{Performance of Code Generation Models on Vulnerability Repair.}
  \label{tab:model_performance}
  \centering
  \scriptsize
  \begin{tabular}{lcc}
    \toprule
    \textbf{Model Name} & \textbf{Vuln. Fixed} & \textbf{Avg. Time (s)} \\
    \midrule
    CodeGeex-4      & 11            & 95.50    \\
    \textbf{CodeGemma}       & \textbf{16}   & 96.50    \\
    CodeLlama       & 3             & 243.93   \\
    CodeQwen        & 5             & 141.05   \\
    CodeStral       & 13            & 295.23   \\
    DeepSeekCoder-33b & 1           & 411.75   \\
    \textbf{DeepSeek-V2}     & 9             & \textbf{19.42}    \\
    Gemma2-9b       & 13            & 108.30   \\
    Gemma2-27b      & 14            & 304.27   \\
    Granite-Code    & 14            & 90.37    \\
    Llama3.2        & 1             & 37.09    \\
    Llama3.3-70b    & 13            & 741.10   \\
    OpenCoder$^*$   & 1$^*$         & 94$^*$   \\
    Qwen-2.5-Code   & 13            & 79.72    \\
    StarCoder2      & 0$^*$         & 89.10    \\
    \textbf{Smartify (Gemma2+CodeGemma)} & \textbf{16} & 112.30   \\
    Smartify (Gemma2+Llama3.1)  & 14      & 267.80   \\
    \bottomrule
  \end{tabular}
\end{table}

% The results reveal significant performance disparities among the evaluated models. \textbf{CodeGemma} emerges as a top performer, successfully fixing 16 vulnerabilities with a relatively low average inference time of 96.5 seconds. This suggests that CodeGemma possesses a strong ability to understand and rectify code vulnerabilities while maintaining reasonable efficiency. Our proposed framework, \textbf{Smartify (Gemma2+CodeGemma)}, achieves comparable performance, also fixing 16 vulnerabilities, albeit with a slightly higher average inference time of 112.3 seconds, likely due to its iterative multi-agent process. \textbf{Gemma2 9b} and \textbf{Gemma2 27b} also demonstrate strong capabilities, fixing 13 and 14 vulnerabilities, respectively. However, the larger Gemma2 27b model exhibits a significantly higher inference time (304.27 seconds) compared to the 9b variant (108.3 seconds), highlighting the trade-off between model size and efficiency. \textbf{Granite-code} performs well, fixing 14 vulnerabilities with an inference time of 90.37 seconds.

The results reveal significant performance disparities among the evaluated models. Among the pre-trained models for Solidity \textbf{CodeGemma} surprisingly emerges as a top performer, successfully fixing 16 vulnerabilities with a relatively low average inference time of 96.5 seconds. This suggests that \textbf{CodeGemma} possesses a strong ability to understand and rectify code vulnerabilities while maintaining reasonable efficiency. However since most of these Solidity smart contracts were part of open Githubs repositories, there can be a strong possibility fo these already being part of the pertaining data. Our proposed framework, \textbf{\sln{} (Gemma2+CodeGemma)}, achieves comparable performance, also fixing 16 vulnerabilities, albeit with a slightly higher average inference time of 112.3 seconds. This increased time is likely due to its iterative multi-agent process, which enables \sln{} to leverage the complementary strengths of \textbf{Gemma2} and \textbf{CodeGemma}, resulting in robust and reliable fixes. 

While \sln{} does not immediately show any benefits over \textbf{CodeGemma} here, we can notice that the same \sln{} framework when applied to \textbf{Llama3.1} without any fine-tuning (unlike the \sln{} with \textbf{CodeGemma}) still gives considerable performance boost over vanilla.

% \textbf{Gemma2 9b} and \textbf{Gemma2 27b} also demonstrate strong capabilities, fixing 13 and 14 vulnerabilities, respectively. However, the larger Gemma2 27b model exhibits a significantly higher inference time (304.27 seconds) compared to the 9b variant (108.3 seconds), highlighting the trade-off between model size and efficiency. \textbf{Granite-code} performs well, fixing 14 vulnerabilities with an inference time of 90.37 seconds, showcasing its competitive performance. 

\begin{figure}
    \centering
    \includegraphics[width=1\linewidth,height=6.2cm]{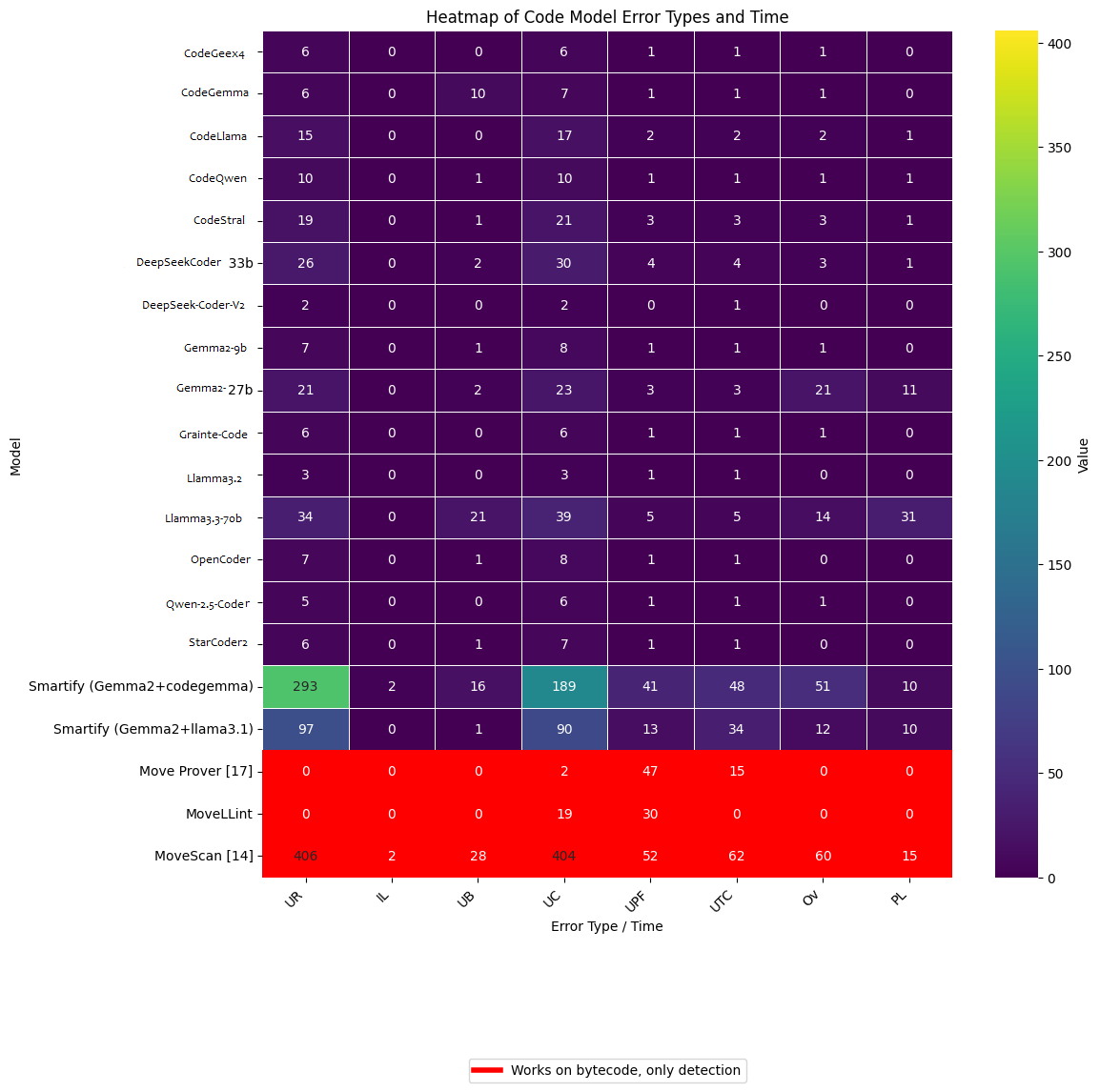}
    \caption{\texttt{Move} Code Repair.}
    \label{fig:enter-label}
\end{figure}

Conversely, models like \textbf{CodeLlama}, \textbf{CodeQWen}, \textbf{DeepSeekCoder 33b}, and \textbf{Llama3.2} show limited effectiveness, fixing only a small number of vulnerabilities. The poor performance of these models could be attributed to several factors, such as insufficient exposure to \texttt{Solidity} code during pre-training or fine-tuning or architectures ill-suited for vulnerability repair, which requires a deep understanding of code syntax and security principles. The exceptionally poor performance of models like \textbf{starcoder2} (marked with an asterisk *), along with incomplete data for \textbf{opencoder}, suggests potential issues with their training data or a fundamental mismatch between their capabilities and the task's demands.  These models might have been trained on an older version of \texttt{Solidity} or different smart contract security practices than those in the Not-So-Smart-Contracts dataset. Moreover, they might prioritize other aspects of code generation, such as code completion, over security-specific tasks like vulnerability repair.

%\vspace{-0.6cm}
\begin{table}[!ht]
\centering
\caption{\texttt{Move} Vulnerability Repair (Time in seconds).}
\label{tab:vulnerability_repair}
%\footnotesize
\scriptsize
\resizebox{\columnwidth}{!}{%
\begin{tabular}{l|c|c|c|c|c|c|c|c|c}
\toprule
Model & \rotatebox{90}{UR} & \rotatebox{90}{IL} & \rotatebox{90}{UB} & \rotatebox{90}{UC} & \rotatebox{90}{UPF} & \rotatebox{90}{UTC} & \rotatebox{90}{Ov} & \rotatebox{90}{PL} & \rotatebox{90}{Time} \\
\midrule
CodeGeex4 & 6 & 0 & 0 & 6 & 1 & 1 & 1 & 0 & 96 \\
CodeGemma & 6 & 0 & 10 & 7 & 1 & 1 & 0 & 0 & 97 \\
CodeLlama & 15 & 0 & 1 & 17 & 2 & 2 & 2 & 1 & 244 \\
CodeQwen & 10 & 0 & 1 & 10 & 1 & 1 & 1 & 1 & 141 \\
CodeStral & 19 & 0 & 1 & 21 & 3 & 3 & 2 & 1 & 295 \\
DeepSeekCoder 33b & 26 & 0 & 2 & 30 & 4 & 4 & 3 & 1 & 412 \\
DeepSeekV2 & 2 & 0 & 0 & 2 & 0 & 1 & 0 & 0 & 19 \\
Gemma2 9b & 7 & 0 & 1 & 8 & 1 & 1 & 0 & 10 & 108 \\
Gemma2 27b & 21 & 0 & 2 & 23 & 3 & 3 & 21 & 11 & 304 \\
Granite-Code & 6 & 0 & 0 & 6 & 1 & 1 & 1 & 0 & 90 \\
Llama3.2 & 3 & 0 & 0 & 3 & 1 & 1 & 0 & 0 & 37 \\
Llama3.3 70b & 34 & 0 & 21 & 39 & 5 & 5 & 14 & 31 & 741 \\
OpenCoder & 7 & 0 & 1 & 8 & 1 & 1 & 0 & 0 & 94* \\
Qwen 2.5 code & 5 & 0 & 0 & 6 & 1 & 1 & 1 & 0 & 80 \\
StarCoder2 & 6 & 0 & 1 & 7 & 1 & 1 & 0 & 0 & 89 \\
    \textbf{Smartify (Gemma2+CodeGemma)} & \textbf{293} & \textbf{2} & \textbf{16} & \textbf{189} & \textbf{41} & \textbf{48} & \textbf{51} & \textbf{10} & \textbf{112} \\
\textbf{Smartify (Gemma2+Llama3.1)} & 97 & 0 & 1 & 90 & 13 & 34 & 12 & \textbf{10} & 268 \\
Move Prover~\cite{dill2022fast} & - & 2 & - & - & - & - & 47 & 15 & - \\
MoveLint & - & - & - & - & 19 & 30 & 0 & 0 & -\\
MoveScan~\cite{song2024empirical} & 406 & 2 & 28 & 404 & 52 & 62 & 60 & 15 & -\\
\bottomrule
\end{tabular}
}
{\scriptsize
\textbf{Abbreviations:} \textbf{UR}: Unchecked Return; \textbf{IL}: Infinite Loop; \textbf{UB}: Unnecessary Boolean; \textbf{UC}: Unused Constant; \textbf{UPF}: Unused Private Function; \textbf{UTC}: Unnecessary Type Conversion; \textbf{Ov}: Overflow; \textbf{PL}: Precision Loss.}
\vspace{-0.8cm}
\end{table}

The public availability of the "Not So Smart Contracts" dataset raises the question of data contamination. Many evaluated models, especially those trained on large, public code corpora, might have encountered this dataset during pre-training, potentially inflating their performance. However, since \textbf{CodeGemma} and \textbf{\sln{} (Gemma2+CodeGemma)} were specifically fine-tuned for this task, the issue of data contamination is likely less significant.

\subsection{\texttt{Move} Code Repair}\label{sec:move_code_repair}

This section analyzes the efficacy of various models in repairing vulnerabilities within \texttt{Move} smart contracts, as detailed in \tablename~\ref{tab:vulnerability_repair}. The evaluation encompasses eight distinct vulnerability categories: {\it Unchecked Return (UR), Infinite Loop (IL), Unnecessary Boolean (UB), Unused Constant (UC), Unused Private Function (UPF), Unnecessary Type Conversion (UTC), Overflow (Ov), and Precision Loss (PL)} following the works of Song et al~\cite{song2024empirical}. The metrics presented in the table represent the number of successfully repaired instances for each vulnerability type, with higher values indicating superior performance. The inference time, measured in seconds, is also provided for each model.

The results demonstrate a significant variance in performance across the evaluated models. Notably, the larger language models, such as \textbf{Deepseekcoder 33b} and \textbf{Llama3.3 70b}, exhibit a relatively higher number of successful repairs across multiple categories, albeit with a corresponding increase in inference time. Conversely, smaller models like \textbf{DeepseekV2} and \textbf{Llama3.2} demonstrate limited repair capabilities.  The specialized tools for \texttt{Move} code, namely \textbf{Move Prover}, \textbf{MoveLint}, and \textbf{MoveScan}, were employed as a benchmark for comparison. It is crucial to note that these tools are designed for vulnerability \textbf{detection} rather than repair. \textbf{MoveScan}, in particular, identified a substantial number of instances across all categories, highlighting its effectiveness as a static analysis tool. \textbf{Move Prover} demonstrated proficiency in detecting Overflow and Precision Loss vulnerabilities, while \textbf{MoveLint} focused on Unused Private Functions and Unnecessary Type Conversions.

The \sln{} models, which take advantage of a combination of \textbf{Gemma2} with either \textbf{CodeGemma} or \textbf{Llama3.1}, present an interesting case. \sln{}(\textbf{Gemma2+CodeGemma}) and \sln{}(\textbf{Gemma2+Llama3.1}) outperform several individual models in multiple categories. This is likely because the specialized models are fine-tuned on the \texttt{Move}-specific dataset. For instance, \sln{} (\textbf{Gemma2+CodeGemma}) achieves the highest number of repairs for the Unchecked Return, Infinite Loop, Unused Boolean, Unused Constant, Unused Private Function, Unnecessary Type Conversion, and Overflow categories, showcasing a substantial improvement over individual models in these areas. However, it is worth mentioning that they also have limitations compared to individual models for certain categories like Precision Loss.

% The results underscore the effectiveness of the \textbf{Smartify} framework in automated Move code vulnerability detection and repair, demonstrating the potential of model combination to enhance performance. It showcases the effectiveness of the proposed architecture. 
%\vspace{0.2cm}
%\underline{This answers our first two research questions.}

\begin{tcolorbox}[
  colback=green!15, % Light green background
  colframe=green!40, % Medium green border
  title=RQ1 \& RQ2 - Code Understanding and Vuln. Detection,
  coltitle=black, % Set title color to black
  fonttitle=\bfseries,
  boxrule=0.75mm, % Thicker border for emphasis
  rounded corners, % Smooth corners for aesthetics
  left=1mm, % Small padding on the left
  right=1mm, % Small padding on the right
  top=1mm, % Small padding on the top
  bottom=1mm % Small padding on the bottom
]
{\footnotesize
\textbf{Yes.} Our empirical analysis with \sln{}, especially with using a fine-tuned Code-Gemma and also using vanilla pre-trained \textbf{Llama3.1}, has shown us the effectiveness of the framework's ability to understand code and capture bad practices leading to vulnerability. Especially for a low-resource code like \texttt{Move}, without significant fine-tuning (in the case of \textbf{Llama3.1}), \sln{} outperform the larger and computationally intensive models such as Llama 3.3 70b. }
\end{tcolorbox}

Notably, \textbf{\sln{} (Gemma2+CodeGemma)}, combining fine-tuned \textbf{Gemma2} with \textbf{CodeGemma}, achieves performance on par with the best individual model, \textbf{CodeGemma}, which is expected due to one of the models being fine-tuned. This highlights the advantages of strategically combining specialized models, answering our next research question.

\begin{tcolorbox}[
  colback=green!15, % Light gray background
  colframe=green!40, % Medium gray border
  title=RQ3 \& RQ4 - Code Repair,
  fonttitle=\bfseries,
    coltitle=black, % Set title color to black
  boxrule=0.75mm, % Thicker border for emphasis
  rounded corners, % Smooth corners for aesthetics
  left=1mm, % Small padding on the left
  right=1mm, % Small padding on the right
  top=1mm, % Small padding on the top
  bottom=1mm % Small padding on the bottom
]
{\footnotesize
Both for \texttt{Solidity} and \texttt{Move}, we were able to compare the efficacy of our framework with prior works. We can see \sln{} outperforms all of the existing code models, even very specialized code models trained on \texttt{Move} (OpenCoder~\cite{huang2024opencoder}) in generating repair codes for detected vulnerabilities.} 
\end{tcolorbox}

Furthermore, \sln{}'s efficacy extends even when integrating a non-finetuned model like \textbf{Llama 3.1}. \sln{} significantly outperforms \textbf{Llama 3.2} by fixing 14 vulnerabilities compared to Llama 3.2's single fix, making its performance comparable with the much more extensive and computationally intensive \textbf{Llama 3.3 70b}. This demonstrates that \sln{}'s architecture can enhance even general-purpose language models for code repair, balancing speed, and accuracy. %Answering our last query:

\begin{tcolorbox}[
  colback=green!15, % Light gray background
  colframe=green!40, % Medium gray border
  title=RQ5 - Generalization,
  fonttitle=\bfseries,
    coltitle=black, % Set title color to black
  boxrule=0.75mm, % Thicker border for emphasis
  rounded corners, % Smooth corners for aesthetics
  left=1mm, % Small padding on the left
  right=1mm, % Small padding on the right
  top=1mm, % Small padding on the top
  bottom=1mm % Small padding on the bottom
]
{\footnotesize
Our implementation of \sln{} with both fine-tuned \textbf{Code-Gemma} and \textbf{Llama3.1} as the second agent allowed us to run our experiments on both sets of LLMs. The results show that \sln{} can significantly boost performance even on non-finetuned models compared to a single model.}
\end{tcolorbox}

Comparative analysis reveals trade-offs between model scale and performance in automated code repair. Larger models, such as \textbf{DeepSeekCoder 33b} and \textbf{Llama 3.3 70b}, exhibit broader repair capabilities but incur higher computational costs and inference times. Conversely, the \textbf{Gemma2 27b} model demonstrates notable proficiency in addressing Overflow vulnerabilities, albeit with limitations in handling Unnecessary Boolean and Unused Constant compared to \textbf{Llama 3.3 70b}. While \textbf{Llama 3.3 70b} outperforms \sln{} in overall repair capability, its significantly slower inference speed poses a challenge for practical deployment. Therefore, for real-world, on-device applications, \textbf{\sln{} (Gemma2+CodeGemma)} presents a compelling solution with its balance of substantial accuracy and rapid inference.

\begin{center}
\fcolorbox{blue!20}{white!90!blue}{%  Box with specified colors
{%\small
    \parbox{\dimexpr\linewidth-2\fboxsep-2\fboxrule}{ % Adjust width for box borders
\textbf{Insight:} Specialized code models like \textbf{StarCoder}~\cite{lozhkov2024starcoder}, \textbf{OpenCoder}~\cite{huang2024opencoder} and \textbf{DeepSeekCoder}~\cite{guo2024deepseek} doesn't necessarily work well even if it's a coding specific task. While code models like \textbf{CodeGemma}~\cite{team2024codegemma} and \textbf{CodeLlama}~\cite{roziere2023code} are much better at understanding instructions and working on code. This helped \sln{} for its understanding and fine-tuning for code repairability.}
    }
}
\end{center}

Specialized static analysis tools for \texttt{Move}, including \textbf{Move Prover}, \textbf{MoveLint}, and \textbf{MoveScan}, work as baselines of detecting \texttt{Move} vulnerabilities with which we compare our \sln{} and other LLMs. These findings underscore the need for targeted model improvements. The \sln{} framework directly addresses these deficiencies, offering enhanced vulnerability repair effectiveness. 

This research also opens up future research directions on using this framework for context-aware test case generation.

\section{Ablation Study}

To rigorously validate our architectural design and isolate the contribution of each key component, a comprehensive ablation study was conducted. This study systematically deconstructs the Smartify framework to quantify the impact of its core mechanisms—agent specialization, iterative refinement, and retrieval-augmented generation—on overall repair performance \cite{huang2025opencoderopencookbooktoptier, jiang2018contractfuzzer}.

\vspace{0.3cm}
\subsection{Ablation Configurations}
\vspace{0.1cm}
Four distinct configurations of the system were evaluated against the Solidity and Move datasets:

\begin{itemize}[leftmargin=*]
    \item \textbf{Full Smartify:} The complete five-agent system, including the RAG module for the Code Generator and the Refiner-Validator iterative feedback loop. This represents our proposed approach.
    \item \textbf{Smartify (No Refinement):} A version of the framework where the Refiner and Validator agents are disabled. The output is taken directly from the first pass of the Code Generator. This configuration measures the impact of the iterative self-improvement loop on patch quality.
    \item \textbf{Smartify (No RAG):} In this setup, the Code Generator operates without the contextual, few-shot examples provided by the RAG system. It relies solely on its fine-tuned knowledge and the Architect's plan. This configuration is designed to measure the importance of providing language-specific, in-context examples, especially for the low-resource Move language.
    \item \textbf{Single-Agent Baseline (CoT-style):} This configuration replaces the multi-agent system with a single, powerful agent (Gemma2 9B). The agent is given a complex, chained prompt that instructs it to sequentially perform the tasks of auditing, planning, and generating the repair. This mimics the linear reasoning workflow of CoT-based approaches like \texttt{ContractTinker} and directly tests the value of our role-based specialization and delegation architecture.
\end{itemize}

\vspace{0.3cm}
\subsection{Analysis}
\vspace{0.1cm}

The performance of each configuration was measured using the Pass@1 and Exploit Mitigation Rate metrics. The results, presented in Table~\ref{tab:ablation_results}, provide clear empirical evidence supporting our architectural choices.

\vspace{0.5cm}
\begin{table}[!htbp]
\centering
\caption{Ablation Study of Smartify Components on Solidity and Move Datasets}
\label{tab:ablation_results}
\resizebox{\linewidth}{!}{%
\begin{tabular}{lcc|cc}
\toprule
 & \multicolumn{2}{c|}{Solidity} & \multicolumn{2}{c}{Move} \\
\cmidrule(lr){2-3} \cmidrule(lr){4-5}
Configuration & Pass@1 (\%) & Exploit Mit. (\%) & Pass@1 (\%) & Exploit Mit. (\%) \\
\midrule
\textbf{Full Smartify} & \textbf{26.7} & \textbf{25.0} & \textbf{48.9} & \textbf{45.7} \\
Smartify (No Refinement) & 25.0 & 20.0 & 44.6 & 39.1 \\
Smartify (No RAG) & 21.7 & 18.3 & 28.3 & 23.9 \\
Single-Agent Baseline & 18.3 & 15.0 & 21.7 & 17.4 \\
\bottomrule
\end{tabular}
}
\end{table}

The results demonstrate a clear and consistent degradation in performance as key components are removed, confirming the positive contribution of each element of the Smartify architecture.

\begin{itemize}
    \item \textbf{Impact of Iterative Refinement:} Removing the refinement loop (\textbf{No Refinement}) causes a noticeable drop in both metrics, particularly the Exploit Mitigation Rate (from 25.0\% to 20.0\% for Solidity). This indicates that while the initial code generation is often correct, the refinement process is crucial for hardening the patch against exploits and catching subtle regressions that the first pass might miss.

    \item \textbf{Impact of RAG:} The removal of the RAG module (\textbf{No RAG}) has a significant negative impact across the board, but its effect is most pronounced for the Move language. The Pass@1 score for Move plummets from 48.9\% to 28.3\%, a relative decrease of over 42\%. This strongly supports our hypothesis that providing in-context, language-specific examples is critical for achieving high performance in low-resource languages where the model's pre-trained knowledge is limited.

    \item \textbf{Impact of Multi-Agent Architecture:} The \textbf{Single-Agent Baseline}, designed to mimic a CoT-style approach, performs the worst of all configurations. Its performance is substantially lower than the Full Smartify framework, particularly on the more complex task of exploit mitigation. This finding provides strong evidence that our role-based, delegative architecture is superior to a monolithic, linear reasoning process. By specializing agents for distinct tasks (analysis vs. planning vs. generation), the system achieves a more robust and effective decomposition of the complex program repair problem.
\end{itemize}

The ablation study empirically validates the design of Smartify. The multi-agent architecture provides a superior structure for complex problem-solving, the refinement loop is essential for ensuring patch quality and security, and the RAG mechanism is a critical component for adapting the framework to new or low-resource programming languages.

\vspace{0.7cm}
\section{Conclusion}
\vspace{0.2cm}

This work addresses the pressing need for enhanced security in the burgeoning blockchain ecosystem. We investigate the application of Large Language Models (LLMs) to smart contract vulnerability detection and repair, focusing on \texttt{Solidity} and \texttt{Move}. 
We introduce \textbf{\sln{}}, a novel multi-agent framework that significantly improves LLM performance in this critical domain. The contributions of this work are: (1) \textbf{\sln{}}, a novel multi-agent framework that enhances LLM-based smart contract vulnerability detection and repair; (2) a method for encoding language-specific knowledge, valuable for low-resource languages like Move; (3) a scalable, adaptable approach applicable to other programming languages and LLMs; (4) a demonstration of \sln{}’s efficacy on generalized pre-trained LLMs; and (5) a detailed analysis of the challenges inherent in automated code repair.

\textbf{\sln{}} represents a significant advancement in automating smart contract security, a crucial concern in the expanding blockchain landscape. Future work will refine the framework, expand its language coverage, particularly within the blockchain domain, and integrate it into real-world blockchain development workflows. This research lays the foundation for AI-powered tools that can bolster the security and reliability of decentralized applications, fostering a more robust and trustworthy blockchain ecosystem.

\section*{Acknowledgments}

This work was partially supported by the Sui Academic Research Award. We gratefully acknowledge the Google Developer Expert AI/ML team for providing cloud compute resources that facilitated this research.

We thank Sam Blackshear of Mysten Labs for his valuable feedback on the idea and early drafts of this paper.

\bibliographystyle{plainnat}
\bibliography{ref}

@inproceedings{nath2014web,
  title={Web 1.0 to Web 3.0-Evolution of the Web and its various challenges},
  author={Nath, Keshab and Dhar, Sourish and Basishtha, Subhash},
  booktitle={2014 International Conference on Reliability Optimization and Information Technology (ICROIT)},
  pages={86--89},
  year={2014},
  organization={IEEE}
}

@article{ray2023web3,
  title={Web3: A comprehensive review on background, technologies, applications, zero-trust architectures, challenges and future directions},
  author={Ray, Partha Pratim},
  journal={Internet of Things and Cyber-Physical Systems},
  volume={3},
  pages={213--248},
  year={2023},
  publisher={Elsevier}
}

@inproceedings{karanjai2021conditional,
  title={On conditional cryptocurrency with privacy},
  author={Karanjai, Rabimba and Xu, Lei and Gao, Zhimin and Chen, Lin and Kaleem, Mudabbir and Shi, Weidong},
  booktitle={2021 IEEE International Conference on Blockchain and Cryptocurrency (ICBC)},
  pages={1--3},
  year={2021},
  organization={IEEE}
}

@misc{huang2024opencoder,
  author       = {Siming Huang and Qingyue Zhang and Yilun Jin and Xiaozhu Meng and Yiyang Wang and Haoyu Wang and Ding Li and Shiqing Ma},
  title        = {Opencoder: The Open Cookbook for Top-Tier Code Large Language Models},
  year         = {2024},
  archivePrefix= {arXiv},
  eprint       = {2411.04905},
  note         = {arXiv preprint},
  url          = {https://arxiv.org/abs/2411.04905}
}

@inproceedings{jiang2018contractfuzzer,
author = {Jiang, Bo and Liu, Ye and Chan, W. K.},
title = {ContractFuzzer: fuzzing smart contracts for vulnerability detection},
year = {2018},
isbn = {9781450359375},
publisher = {Association for Computing Machinery},
address = {New York, NY, USA},
url = {https://doi.org/10.1145/3238147.3238177},
doi = {10.1145/3238147.3238177},
booktitle = {Proceedings of the 33rd ACM/IEEE International Conference on Automated Software Engineering},
pages = {259–269},
numpages = {11},
location = {Montpellier, France},
series = {ASE '18}
}

@misc{huang2025opencoderopencookbooktoptier,
      title={OpenCoder: The Open Cookbook for Top-Tier Code Large Language Models}, 
      author={Siming Huang and Tianhao Cheng and J. K. Liu and Jiaran Hao and Liuyihan Song and Yang Xu and J. Yang and Jiaheng Liu and Chenchen Zhang and Linzheng Chai and Ruifeng Yuan and Zhaoxiang Zhang and Jie Fu and Qian Liu and Ge Zhang and Zili Wang and Yuan Qi and Yinghui Xu and Wei Chu},
      year={2025},
      eprint={2411.04905},
      archivePrefix={arXiv},
      primaryClass={cs.CL},
      url={https://arxiv.org/abs/2411.04905}, 
}

@inproceedings{wang2024contracttinker,
  title={Contracttinker: Llm-empowered vulnerability repair for real-world smart contracts},
  author={Wang, Che and Zhang, Jiashuo and Gao, Jianbo and Xia, Libin and Guan, Zhi and Chen, Zhong},
  booktitle={Proceedings of the 39th IEEE/ACM International Conference on Automated Software Engineering},
  pages={2350--2353},
  year={2024}
}

@misc{bobadilla2025automated,
  author       = {Sofia Bobadilla and Monica Jin and Martin Monperrus},
  title        = {Do Automated Fixes Truly Mitigate Smart Contract Exploits?},
  year         = {2025},
  archivePrefix= {arXiv},
  eprint       = {2501.04600},
  note         = {arXiv e-prints},
  url          = {https://arxiv.org/abs/2501.04600}
}

@inproceedings{grieco2020echidna,
  author    = {Gustavo Grieco and Santiago Palladino and Martin Ortigoza and Federico Bond and Valentin Wustholz},
  title     = {Echidna: Effective, Usable, and Fast Fuzzing for Smart Contracts},
  booktitle = {Proceedings of the 29th ACM SIGSOFT International Symposium on Software Testing and Analysis (ISSTA '20)},
  year      = {2020},
  pages     = {481--493},
  publisher = {ACM},
  doi       = {10.1145/3395363.3397384}
}

@inproceedings{kaleem2021event,
  title={An event driven framework for smart contract execution},
  author={Kaleem, Mudabbir and Kasichainula, Keshav and Karanjai, Rabimba and Xu, Lei and Gao, Zhimin and Chen, Lin and Shi, Weidong},
  booktitle={Proceedings of the 15th ACM International Conference on Distributed and Event-based Systems},
  pages={78--89},
  year={2021}
}

@article{zheng2018blockchain,
  title={Blockchain challenges and opportunities: A survey},
  author={Zheng, Zibin and Xie, Shaoan and Dai, Hong-Ning and Chen, Xiangping and Wang, Huaimin},
  journal={International journal of web and grid services},
  volume={14},
  number={4},
  pages={352--375},
  year={2018},
  publisher={Inderscience Publishers (IEL)}
}

@article{vacca2021systematic,
  title={A systematic literature review of blockchain and smart contract development: Techniques, tools, and open challenges},
  author={Vacca, Anna and Di Sorbo, Andrea and Visaggio, Corrado A and Canfora, Gerardo},
  journal={Journal of Systems and Software},
  volume={174},
  pages={110891},
  year={2021},
  publisher={Elsevier}
}

@article{li2023smart,
  title={A Smart Contract Vulnerability Detection Method Based on Multimodal Feature Fusion and Deep Learning},
  author={Li, Jinggang and Lu, Gehao and Gao, Yulian and Gao, Feng},
  journal={Mathematics},
  volume={11},
  number={23},
  pages={4823},
  year={2023},
  publisher={MDPI}
}

@article{wronka2023financial,
  title={Financial crime in the decentralized finance ecosystem: new challenges for compliance},
  author={Wronka, Christoph},
  journal={Journal of Financial Crime},
  volume={30},
  number={1},
  pages={97--113},
  year={2023},
  publisher={Emerald Publishing Limited}
}

@article{popper2016hacking,
  title={A hacking of more than \$50 million dashes hopes in the world of virtual currency},
  author={Popper, Nathaniel},
  journal={The New York Times},
  volume={17},
  year={2016}
}

@Online {polyhack,
  title = {Decoding Poly Network \$34 Billion Hack},
  date = {2025-01-02},
  year = {2025},
  url = {https://www.quillaudits.com/blog/hack-analysis/poly-network-hack},
  urldate = {2025-01-02}
}

@inproceedings{joshi2023repair,
  title={Repair is nearly generation: Multilingual program repair with llms},
  author={Joshi, Harshit and Sanchez, Jos{\'e} Cambronero and Gulwani, Sumit and Le, Vu and Verbruggen, Gust and Radi{\v{c}}ek, Ivan},
  booktitle={Proceedings of the AAAI Conference on Artificial Intelligence},
  volume={37},
  number={4},
  pages={5131--5140},
  year={2023}
}

@inproceedings{10.1145/3238147.3238177,
author = {Jiang, Bo and Liu, Ye and Chan, W. K.},
title = {ContractFuzzer: fuzzing smart contracts for vulnerability detection},
year = {2018},
isbn = {9781450359375},
publisher = {Association for Computing Machinery},
address = {New York, NY, USA},
url = {https://doi.org/10.1145/3238147.3238177},
doi = {10.1145/3238147.3238177},
booktitle = {Proceedings of the 33rd ACM/IEEE International Conference on Automated Software Engineering},
pages = {259–269},
numpages = {11},
location = {Montpellier, France},
series = {ASE '18}
}

@article{wang2020contractward,
  title={Contractward: Automated vulnerability detection models for ethereum smart contracts},
  author={Wang, Wei and Song, Jingjing and Xu, Guangquan and Li, Yidong and Wang, Hao and Su, Chunhua},
  journal={IEEE Transactions on Network Science and Engineering},
  volume={8},
  number={2},
  pages={1133--1144},
  year={2020},
  publisher={IEEE}
}

@inproceedings{song2024empirical,
  title={Empirical Study of Move Smart Contract Security: Introducing MoveScan for Enhanced Analysis},
  author={Song, Shuwei and Chen, Jiachi and Chen, Ting and Luo, Xiapu and Li, Teng and Yang, Wenwu and Wang, Leqing and Zhang, Weijie and Luo, Feng and He, Zheyuan and others},
  booktitle={Proceedings of the 33rd ACM SIGSOFT International Symposium on Software Testing and Analysis},
  pages={1682--1694},
  year={2024}
}

@article{blackshear2019move,
  title={Move: A language with programmable resources},
  author={Blackshear, Sam and Cheng, Evan and Dill, David L and Gao, Victor and Maurer, Ben and Nowacki, Todd and Pott, Alistair and Qadeer, Shaz and Rain, Dario Russi and Sezer, Stephane and others},
  journal={Libra Assoc},
  pages={1},
  year={2019}
}

@article{blackshear2022move,
  title={The Move Borrow Checker},
  author={Blackshear, Sam and Mitchell, John and Nowacki, Todd and Qadeer, Shaz},
  journal={arXiv preprint arXiv:2205.05181},
  year={2022}
}

@inproceedings{dill2022fast,
  title={Fast and reliable formal verification of smart contracts with the move prover},
  author={Dill, David and Grieskamp, Wolfgang and Park, Junkil and Qadeer, Shaz and Xu, Meng and Zhong, Emma},
  booktitle={International Conference on Tools and Algorithms for the Construction and Analysis of Systems},
  pages={183--200},
  year={2022},
  organization={Springer}
}

@inproceedings{keilty2022model,
  title={A model-checking framework for the verification of move smart contracts},
  author={Keilty, Eric and Nelaturu, Keerthi and Wu, Bowen and Veneris, Andreas},
  booktitle={2022 IEEE 13th International Conference on Software Engineering and Service Science (ICSESS)},
  pages={1--7},
  year={2022},
  organization={IEEE}
}

@inproceedings{park2024securing,
  title={Securing Aptos framework with formal verification},
  author={Park, Junkil and Zhang, Teng and Grieskamp, Wolfgang and Xu, Meng and Di Giacomo, Gerardo and Chen, Kundu and Lu, Yi and Chen, Robert},
  booktitle={5th International Workshop on Formal Methods for Blockchains (FMBC 2024)},
  year={2024},
  organization={Schloss Dagstuhl--Leibniz-Zentrum f{\"u}r Informatik}
}

@misc{githubGitHubCryticnotsosmartcontracts,
	author = {},
	title = {not-so-smart-contracts: {E}xamples of {S}olidity security issues},
	howpublished = {\url{https://github.com/crytic/not-so-smart-contracts}},
	year = {},
	note = {[Accessed 09-01-2025]},
}

@misc{huggingfaceCodeModels,
	author = {},
	title = {{B}ig {C}ode {M}odels {L}eaderboard - a {H}ugging {F}ace {S}pace by bigcode --- huggingface.co},
	howpublished = {\url{https://huggingface.co/spaces/bigcode/bigcode-models-leaderboard}},
	year = {},
	note = {[Accessed 10-01-2025]},
}

@inproceedings{feist2019slither,
  title={Slither: a static analysis framework for smart contracts},
  author={Feist, Josselin and Grieco, Gustavo and Groce, Alex},
  booktitle={2019 IEEE/ACM 2nd International Workshop on Emerging Trends in Software Engineering for Blockchain (WETSEB)},
  pages={8--15},
  year={2019},
  organization={IEEE}
}

@article{muellerfile,
  title={File 1 of},
  author={Mueller, Bernhard}
}

@inproceedings{mossberg2019manticore,
  title={Manticore: A user-friendly symbolic execution framework for binaries and smart contracts},
  author={Mossberg, Mark and Manzano, Felipe and Hennenfent, Eric and Groce, Alex and Grieco, Gustavo and Feist, Josselin and Brunson, Trent and Dinaburg, Artem},
  booktitle={2019 34th IEEE/ACM International Conference on Automated Software Engineering (ASE)},
  pages={1186--1189},
  year={2019},
  organization={IEEE}
}

@inproceedings{hildenbrandt2018kevm,
  title={Kevm: A complete formal semantics of the ethereum virtual machine},
  author={Hildenbrandt, Everett and Saxena, Manasvi and Rodrigues, Nishant and Zhu, Xiaoran and Daian, Philip and Guth, Dwight and Moore, Brandon and Park, Daejun and Zhang, Yi and Stefanescu, Andrei and others},
  booktitle={2018 IEEE 31st Computer Security Foundations Symposium (CSF)},
  pages={204--217},
  year={2018},
  organization={IEEE}
}

@inproceedings{zhong2020move,
  title={The move prover},
  author={Zhong, Jingyi Emma and Cheang, Kevin and Qadeer, Shaz and Grieskamp, Wolfgang and Blackshear, Sam and Park, Junkil and Zohar, Yoni and Barrett, Clark and Dill, David L},
  booktitle={Computer Aided Verification: 32nd International Conference, CAV 2020, Los Angeles, CA, USA, July 21--24, 2020, Proceedings, Part I 32},
  pages={137--150},
  year={2020},
  organization={Springer}
}

@article{roziere2023code,
  title={Code llama: Open foundation models for code},
  author={Roziere, Baptiste and Gehring, Jonas and Gloeckle, Fabian and Sootla, Sten and Gat, Itai and Tan, Xiaoqing Ellen and Adi, Yossi and Liu, Jingyu and Sauvestre, Romain and Remez, Tal and others},
  journal={arXiv preprint arXiv:2308.12950},
  year={2023}
}

@misc{devaptos,
  title={The aptos blockchain: Safe, scalable, and upgradeable web3 infrastructure},
  author={Dev, Aptos}
}

@inproceedings{blackshear2024sui,
  title={Sui lutris: A blockchain combining broadcast and consensus},
  author={Blackshear, Sam and Chursin, Andrey and Danezis, George and Kichidis, Anastasios and Kokoris-Kogias, Lefteris and Li, Xun and Logan, Mark and Menon, Ashok and Nowacki, Todd and Sonnino, Alberto and others},
  booktitle={Proceedings of the 2024 on ACM SIGSAC Conference on Computer and Communications Security},
  pages={2606--2620},
  year={2024}
}

@Online {starcoin,
  title = {Starcoin Whitepaper},
  date = {2024-11-26},
  year = {2024},
  url = {https://starcoin.org/_astro/whitepaper.CMlZ6t\_x.pdf},
  urldate = {2025-01-10}
}

@article{dannen2017solidity,
  title={Solidity programming},
  author={Dannen, Chris and Dannen, Chris},
  journal={Introducing Ethereum and Solidity: Foundations of Cryptocurrency and Blockchain Programming for Beginners},
  pages={69--88},
  year={2017},
  publisher={Springer}
}

@misc{lozhkov2024starcoder,
      title={StarCoder 2 and The Stack v2: The Next Generation}, 
      author={Anton Lozhkov and Raymond Li and Loubna Ben Allal and Federico Cassano and Joel Lamy-Poirier and Nouamane Tazi and Ao Tang and Dmytro Pykhtar and Jiawei Liu and Yuxiang Wei and Tianyang Liu and Max Tian and Denis Kocetkov and Arthur Zucker and Younes Belkada and Zijian Wang and Qian Liu and Dmitry Abulkhanov and Indraneil Paul and Zhuang Li and Wen-Ding Li and Megan Risdal and Jia Li and Jian Zhu and Terry Yue Zhuo and Evgenii Zheltonozhskii and Nii Osae Osae Dade and Wenhao Yu and Lucas Krauß and Naman Jain and Yixuan Su and Xuanli He and Manan Dey and Edoardo Abati and Yekun Chai and Niklas Muennighoff and Xiangru Tang and Muhtasham Oblokulov and Christopher Akiki and Marc Marone and Chenghao Mou and Mayank Mishra and Alex Gu and Binyuan Hui and Tri Dao and Armel Zebaze and Olivier Dehaene and Nicolas Patry and Canwen Xu and Julian McAuley and Han Hu and Torsten Scholak and Sebastien Paquet and Jennifer Robinson and Carolyn Jane Anderson and Nicolas Chapados and Mostofa Patwary and Nima Tajbakhsh and Yacine Jernite and Carlos Muñoz Ferrandis and Lingming Zhang and Sean Hughes and Thomas Wolf and Arjun Guha and Leandro von Werra and Harm de Vries},
      year={2024},
      eprint={2402.19173},
      archivePrefix={arXiv},
      primaryClass={cs.SE}
}

@article{guo2024deepseek,
  title={DeepSeek-Coder: When the Large Language Model Meets Programming--The Rise of Code Intelligence},
  author={Guo, Daya and Zhu, Qihao and Yang, Dejian and Xie, Zhenda and Dong, Kai and Zhang, Wentao and Chen, Guanting and Bi, Xiao and Wu, Yu and Li, YK and others},
  journal={arXiv preprint arXiv:2401.14196},
  year={2024}
}

@article{team2024codegemma,
  title={Codegemma: Open code models based on gemma},
  author={Team, CodeGemma and Zhao, Heri and Hui, Jeffrey and Howland, Joshua and Nguyen, Nam and Zuo, Siqi and Hu, Andrea and Choquette-Choo, Christopher A and Shen, Jingyue and Kelley, Joe and others},
  journal={arXiv preprint arXiv:2406.11409},
  year={2024}
}
\end{document}